\documentclass[twocolumn,preprintnumbers,floatfix,prb,superscriptaddress]{revtex4}
\usepackage{amsmath}
\usepackage{graphicx}
\usepackage{dcolumn} 
\usepackage{bm}
\usepackage{epsfig}
\usepackage{xcolor}

\begin{document} 
\title{Ab initio study of $Z_2$
topological phases in perovskite (111)
$(\text{SrTiO}_3)_7/(\text{SrIrO}_3)_2$ and
$(\text{KTaO}_3)_7/(\text{KPtO}_3)_2$
multilayers
}

\author{J. L. Lado}
\email{jose.luis.lado@gmail.com}
\affiliation{
Departamento de F\'{\i}sica Aplicada, Universidade
de Santiago de Compostela, E-15782 Santiago de Compostela,
Spain
}
\affiliation{
Instituto de Investigaci\'{o}ns Tecnol\'{o}xicas, Universidade
de Santiago de Compostela, E-15782 Santiago de Compostela,
Spain
}
\author{V. Pardo}
\email{victor.pardo@usc.es}
\affiliation{
Departamento de F\'{\i}sica Aplicada, Universidade
de Santiago de Compostela, E-15782 Santiago de Compostela,
Spain
}
\affiliation{
Instituto de Investigaci\'{o}ns Tecnol\'{o}xicas, Universidade
de Santiago de Compostela, E-15782 Santiago de Compostela,
Spain
}
\author{D. Baldomir}
\affiliation{
Departamento de F\'{\i}sica Aplicada, Universidade
de Santiago de Compostela, E-15782 Santiago de Compostela,
Spain
}
\affiliation{
Instituto de Investigaci\'{o}ns Tecnol\'{o}xicas, Universidade
de Santiago de Compostela, E-15782 Santiago de Compostela,
Spain
}

\date{\today}

\begin{abstract}
Honeycomb structures formed by the growth of perovskite 5d transition metal
oxide
heterostructures
along the (111) direction
in $t_{2g}^5$ configuration can give rise to topological ground states
characterized by a topological index $\nu$=1, as found in Nature Commun. 2,
596 (2011).
Using a combination of a tight binding model and ab initio calculations
we study the multilayers $(\text{SrTiO}_3)_7/(\text{SrIrO}_3)_2$ and 
$(\text{KTaO}_3)_7/(\text{KPtO}_3)_2$
as a function of parity asymmetry, on-site interaction
and uniaxial strain and determine the nature and evolution of the gap.
According to our DFT calculations,
$(\text{SrTiO}_3)_7/(\text{SrIrO}_3)_2$ is found to be a topological
semimetal whereas
$(\text{KTaO}_3)_7/(\text{KPtO}_3)_2$
is found to present a
topological insulating phase that can be understood
as the high U
limit of the previous one,
that can be driven to a trivial insulating phase
by a perpendicular
external electric field.

\end{abstract}

\maketitle

\section{Introduction}
Topological insulators (TI)\cite{topins,reviewti} are a type of
materials which show a gapped
bulk spectrum but gapless surface states.
The topological nature of the surface states
protects them against perturbations and backscattering.
\cite{robust2,robust3,robust5,robust6,robust7,robust8,robust9}
In addition, 
the surface of a d-dimensional TI is such that the effective Hamiltonian
defined on its surface cannot be represented by the Hamiltonian
of a d-1 dimensional
material with the same symmetries,
so the physics of a (d-1)-surface of a topological
insulator
may show completely different behavior from that of a
conventional (d-1)-dimensional material.
Surface
states\cite{bulk-edge}
can be understood in terms of solitonic states which interpolate
between two topologically different vacuums, the topological vacuum of the TI
and the trivial vacuum of a conventional insulator or empty space.

The ground state of a system can be classified by
a certain topological number\cite{top-num,kane-qshe}
depending on its dimensionality
and symmetries present which define its topological classification.
\cite{clas-sup,reviewti}
Two-dimensional single-particle Hamiltonians
with time reversal (TR) invariance are classified
in a $Z_2$ ($\nu=0,1$) topological class.\cite{kane-qshe} Two-dimensional TR systems with
nontrivial topological index ($\nu=1$) show the so-called
quantum spin Hall effect (QSHE) which is characterized by a 
non-vanishing spin Chern number \cite{index-qshe} and a helical edge current.
\cite{current-qshe} This state has been theoretically predicted and
experimentally confirmed in HgTe quantum wells,\cite{hg-te,hg-te2,hg-te3}
as well as predicted in several materials such as
two-dimensional Si and Ge\cite{si-qshe}
and
transition metal oxide (TMO) heterostructures.\cite{xiao}
All these systems have in common that they
present a honeycomb lattice structure with two atoms (A,B) as atomic basis.
In that situation,
it is tempting to think that the effective Hamiltonian in
certain k-points could have the form of a Dirac Hamiltonian.
The components of the spinor 
would be some combination
of localized orbitals in the A or B atoms, whereas
the coupling would take place via non-diagonal elements due to the
bipartite geometry of the lattice.
The simplest and best known
example is graphene, where the Hamiltonian
is a Dirac equation in two nonequivalent points K and K'.
At half filling,
graphene with TR and inversion symmetry (IS)
has $\nu=1$,\cite{prb-kane} so a term which does not break those symmetries
and opens a gap in the whole Brillouin zone would give rise to the QSHE
state.
The way in which an IS term can arise in a graphene Hamiltonian
is due to spin-orbit coupling (SOC), however it is known that the
gap opened this way is
too small.\cite{soc-gra} In contrast, a sublattice asymmetry, which breaks IS,
will open a trivial gap, as in BN.\cite{bn-dft,bn-gra} 

How a honeycomb structure can be constructed from a perovskite unit cell
can be seen in Fig. 
1a and 1b, a perovskite bilayer grown along the (111) direction made of
an open-shell oxide is sandwiched by an isostructural band insulating
oxide. The metal atoms of the bilayer form a buckled honeycomb lattice.
It has been shown 
that perovskite (and also pyrochlore) (111) multilayers 
can develop
topological phases,
\cite{xiao,111-orb,xiao2,ir-dice,la-al,pirocloro,pirocloro2}
as well as
spin-liquid phases and
non-trivial superconducting states.\cite{spin-liq-111}
Topological insulating phases have been predicted for various fillings of
the d shell,\cite{xiao} here we will focus on the large SOC limit
(5d electrons) and formal $d^5$ filling.
We will study two different multilayers,
$(\text{SrTiO}_3)_7/(\text{SrIrO}_3)_2$
and $(\text{KTaO}_3)_7/(\text{KPtO}_3)_2$ 
and we will focus on the realization of a nontrivial $\nu=1$ ground state. 
$\text{SrIrO}_3$ ($a_{\text{SrIrO}_3}=3.94$ \AA)\cite{sriro3} is a correlated
metal\cite{sriro3-elec} whose lattice
match with $\text{SrTiO}_3$
(STO) would be close
enough
($a_{\text{SrTiO}_3}=3.905$ \AA)\cite{srtio3}
for them to grow epitaxially with standard growth techniques.\cite{grow111}
$\text{KPtO}_3$ has not been synthesized (to the best
of our knowledge) but our calculations ($a_{\text{KPtO}_3}=4.02$ \AA)
show a reasonable lattice match with
$\text{KTaO}_3$ would be possible ($a_{\text{KTaO}_3}=3.98$ \AA).\cite{ktao3}
The first multilayer is an iridate
very similar to the well known  
$\text{Na}_2\text{IrO}_3$.\cite{Na2IrO3,arpNa,1evir} This system presents
a layered honeycomb lattice
of Ir atoms at $t_{2g}^5$ filling,
whereas the present bilayers show
a buckled honeycomb lattice, and is predicted
to develop the QSHE, however electron correlation would lead to
an antiferromagnetic order in the edges. 
We will study the dependence
of the topological ground state
on the applied uniaxial strain and the electron-electron
interaction and we will determine a transition between two topological phases
in both materials.
The work is organized as follows. In Section II we introduce a simple
tight binding (TB) model as in Ref. \onlinecite{xiao} focusing on the $t_{2g}^5$ case.
In Section III
we use density functional theory (DFT)
calculations to study the evolution of both multilayers
with uniaxial strain and on-site Coulomb repulsion and we determine the ground state
of each material. In Section IV we study the stability of the topological
phase against TR and IS breaking using both
TB and DFT calculations. Finally in Section V we summarize the results obtained.

\begin{figure}


                \includegraphics[width=\columnwidth]{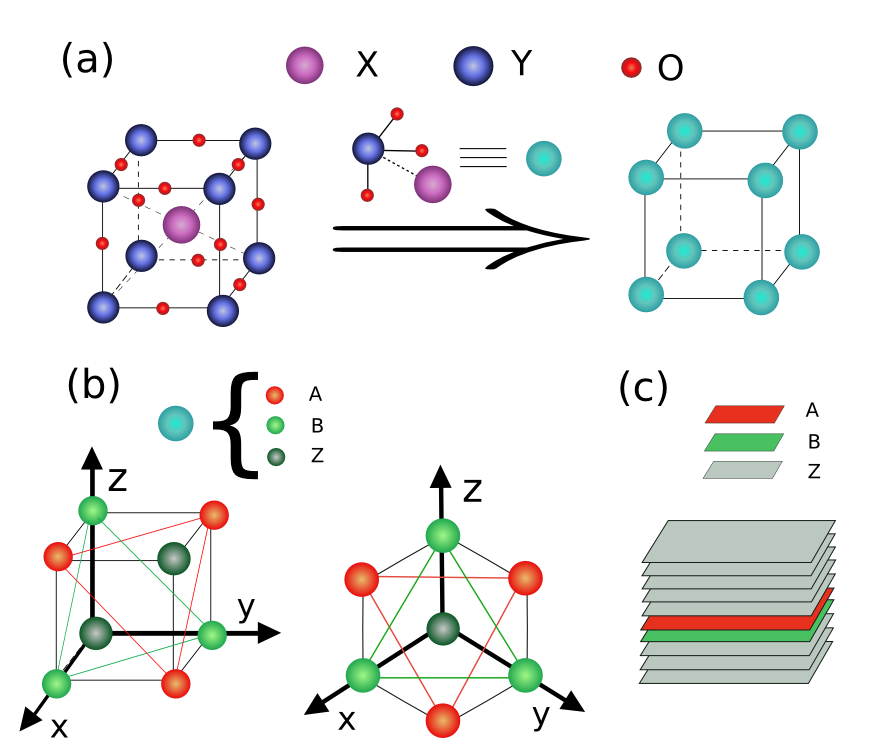}


\caption{(Color online) (a) Scheme of the cubic perovskite structure $XYO_3$.
(b) Construction of the bilayer,
the TM atoms are arranged in a triangular A and B
lattice in the (111) direction in
such a way the two atoms
will form a honeycomb lattice. The Z atom corresponds to the
insulating layer (in our case $\text{SrTiO}_3$ or $\text{KTaO}_3$) and
does not
participate in the honeycomb. (c) Scheme of the multilayer considered in the DFT calculations.}
\label{bilayer}
\end{figure}

\section{Tight binding model}
The qualitative behavior of this system
can be understood using a simple TB
model for the 5d electrons in the TM atoms, as shown
in Ref. \onlinecite{xiao}.
In Section IIA we will give the qualitative behavior
of the effective Hamiltonian.
In Section IIB we will show numerical calculations
of the full model.

\subsection{Full Hamiltonian}
The octahedral environment of oxygen atoms surrounding the transition metal
atoms decouples the d levels in a $t_{2g}$ sextuplet and an $e_g$ quadruplet.
Given that the crystal field gap is higher than the other parameters
considered we will retain only the $t_{2g}$ orbitals.
The Hamiltonian considered for the $t_{2g}$ levels takes the form

\begin{equation}
H=H_{SO}+H_{t}+H_{tri}+H_m
\end{equation}

$H_{SO}$ is the SOC term, which gives rise to an effective
angular
momentum
$J_{eff}=S-L$
which decouples the $t_{2g}$ levels into a filled
j=3/2 quadruplet and a half filled $j=1/2$ doublet.
$H_t$ is the hopping between
neighboring atoms via oxygen that couples the local orbitals.
$H_{tri}$ is a local trigonal term\cite{xiao}
which is responsible for opening
a gap (as we will see below)
without breaking TR and IS. $H_{m}$ is a term which breaks IS 
making the two sublattices nonequivalent tending to open a trivial gap
by decoupling them. In the following discussion we will suppose
that this last term is zero,
but we will analyze its role
in Section IV.

We are interested in two different regimes as a function of SOC strength:
strong and intermediate.
We call strong SOC to the regime where the j=3/2 and j=1/2 are completely
decoupled so that there is a trivial gap between them.
We will refer to an intermediate
regime if the two subsets
are coupled by the hopping.
The key point
to understand the topological character of the calculations is that
a $t_{2g}^5$ configuration can be adiabatically connected
from the strong to the intermediate
regime without closing the gap.    
The argument is the following, beginning in the strong SOC regime it is expected
that a four-band effective model will be a good approximation.
In this regime the mathematical structure of the effective Hamiltonian
turns out to be equivalent to graphene.
The trigonal
term is responsible for opening a gap $\Delta$ via a third order process
in perturbation theory

\begin{equation}
\Delta\sim\lambda_{tri}\frac{t^2}{\alpha^2}
\end{equation} 
where $\lambda_{tri}$ is the trigonal coupling,
$t$ is the hopping parameter and $\alpha$ the SOC strength.
It can be checked by
symmetry considerations that
the restriction of the matrix representations leads
to this term as the first non-vanishing contribution
in perturbation theory. Eq. (2) has
been checked by a logarithmic fitting of numerical
calculations of the full model.
This term will open a gap in the K point conserving TR and IS and thus
realizing a $\nu=1$ ground state.\cite{prb-kane}

As SOC decreases, the gap becomes larger while the system evolves
from the strong to the intermediate regime, so
the intermediate regime is expected to be a topological configuration
\cite{xiao} with
a non-vanishing gap. Note that even though perturbation theory will only hold
in the strong SOC regime, the increase in the gap as the system goes to the
intermediate regime suggests that the $t_{2g}^5$ configuration will
always remain
gapped. This argument is checked by the numerical calculations shown below.

\begin{figure}


                \includegraphics[width=\columnwidth]{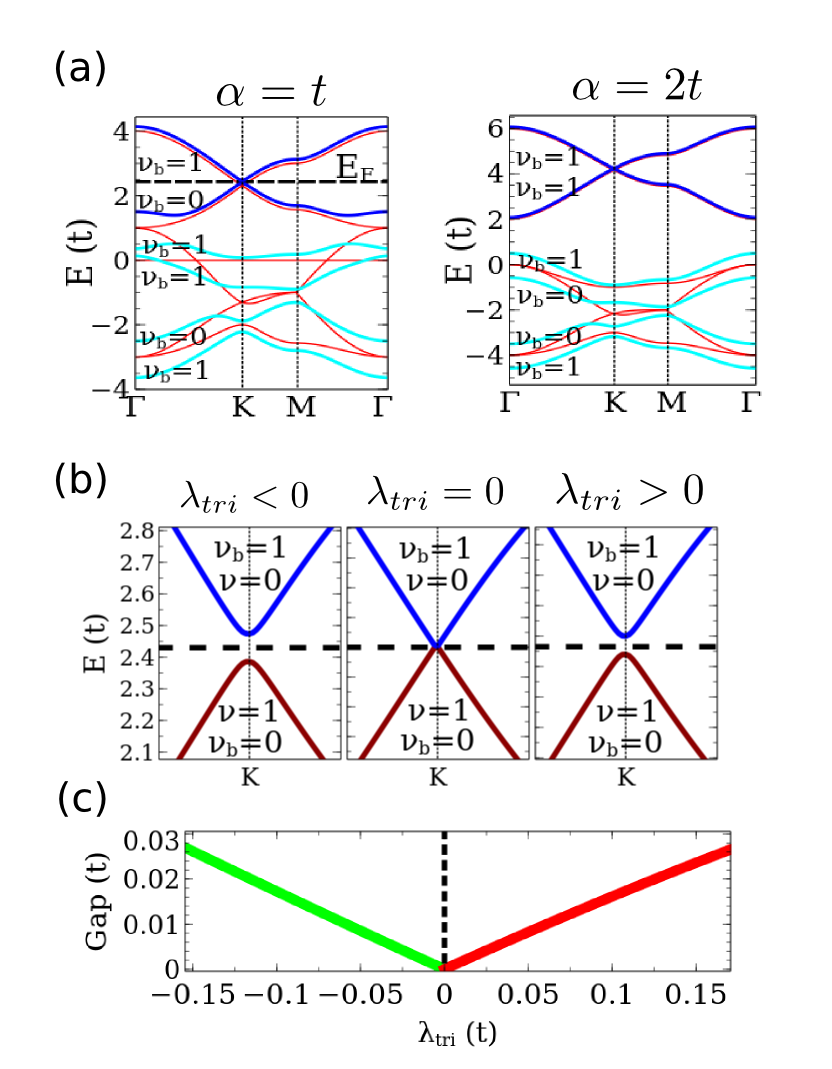}



\caption{(Color online) (a) Band structure of the TB model
with intermediate $\alpha=t$
(left) and strong $\alpha=2t$ (right) SOC strength
and $\lambda_{tri}=-0.5t$.
The difference between the two cases relies on the $\nu_b$ invariant of the last
filled band. The red lines are the band structure with $\lambda_{tri}=0$. (b) Band structure zoomed for the j=1/2 bands near the
K point for negative ($\lambda_{tri}=-0.5t$),
zero and positive ($\lambda_{tri}=0.5t$) trigonal coupling. In the three
cases the topological invariant gives a topological ground state.
(c) Evolution of the gap in the K point with $\lambda_{tri}$ for the
intermediate SOC regime. The two topological phases found will
be identified as HUTI and LUTI in the DFT calculations.}
\label{tight}
\end{figure}

\subsection{Results from tight binding calculations}
In Fig. 2
we show the results of a calculation using
the TB Hamiltonian proposed above. Figure
2a
is the
bulk band structure for strong
and intermediate SOC strength $\alpha$.
If a non-vanishing trigonal term is included, it opens a gap
in the Dirac points of the band structure
generating
topologically non-trivial configurations. We can see this clearly in Fig.
2b, where the band structure close to the Fermi level in the vicinity of the K point is shown.

The topological character of each configuration is defined by the $\nu$
topological invariant which for a band in
an IS Hamiltonian can be calculated as \cite{prb-kane}
\begin{equation}
(-1)^{\nu_b}=\prod_{\text{TRIM}}\langle\Psi_b|P|\Psi_b\rangle
\label{parities}
\end{equation}
where the product runs over the four time reversal invariant momenta (TRIM).
The full invariant of a configuration will be the product of the last
equation over all the occupied bands
\begin{equation}
\nu=\sum_{\text{occ.bands}} \nu_b \text{ (mod 2)}
\label{par-full}
\end{equation}

For a $t_{2g}^5$ filling
the first unfilled band has always
$\nu_b=1$ but the difference between strong and intermediate SOC is the
$\nu_b$ invariant of the last filled band. For strong SOC the j=1/2 and j=3/2
are completely decoupled so a $t_{2g}^4$ configuration
would be topologically trivial,
being the invariant of the fifth band $\nu_b=1$. However, when SOC is not
sizable the bandwidths are large enough to couple the j=1/2 and j=3/2 levels
so that the $t_{2g}^4$ filling is a topological configuration. In both cases
the $t_{2g}^5$ filling is topologically non-trivial. \cite{xiao} We will 
see below using DFT calculations that the systems under study
(TMO's with 5d electrons in a perovskite bilayer structure)
 are in this
intermediate SOC regime.

Figure
2b
shows the bulk band structure, focusing now on the
j=1/2 bands, for negative, zero and positive trigonal terms.
The left numbers are the $\nu_b$ invariant of the
band while the right numbers are the sum of the invariants of that band and the
bands below it.  
No matter what the sign of $\lambda_{tri}$ is,
the configuration becomes non-trivial, being its role
to open a gap in the K point 
around the Fermi level.
In the
DFT calculations below, it will be seen that a change of sign of the trigonal
term can be understood as a topological transition between a 
low U topological
insulating phase (LUTI) and a high U topological insulator (HUTI),
across
a boundary where the system behaves as a topological semimetal (TSM).

\section{DFT calculations}

\subsection{Computational procedures}
Ab initio electronic structure calculations have been performed
using the all-electron full potential code {\sc wien2k} \cite{wien}
The unit cell chosen
is shown in Fig. 1c.
It consists of 9 perovskite layers grown along the (111) direction,
2 layers of
$\text{SrIrO}_3$ ($\text{KPtO}_3$)
which conform the honeycomb and
7 layers of
$\text{SrTiO}_3$ ($\text{KTaO}_3$)
which isolate one honeycomb from the other.

For the different off-plane lattice parameters along the (111) direction
of the perovskite considered,
the structure was relaxed using the full symmetry of the original cell.
The exchange-correlation term is parametrized depending
on the case using the generalized gradient
approximation
(GGA) in the Perdew-Burke-Ernzerhof\cite{gga} scheme,
local density approximation+U
(LDA+U) in the so-called "fully located limit"\cite{sic1} and 
the Tran-Blaha modified Becke-Jonsson (TB-mBJ) potential.\cite{tbmbjlda}

The calculations were performed with a k-mesh of
7 $\times$ 7 $\times$1, a value of  R$_{mt}$K$_{max}$= 7.0.
SOC was introduced in a second variational manner using the scalar
relativistic approximation.\cite{singh}
The $R_{mt}$ values used were in a.u.
1.89 for Ti, 1.91 for Ir, 2.5 for Sr and 1.67 for O in the
$(\text{SrTiO}_3)_7/(\text{SrIrO}_3)_2$
multilayer
and
1.93 for Ta, 1.92 for Pt, 2.5 for K, 1.7 for O
in the
$(\text{KTaO}_3)_7/(\text{KPtO}_3)_2$
multilayer.

\subsection{Band structure of the non-magnetic ground state}

We have already discussed that the systems chosen to study a $d^5$ filling
in a honeycomb lattice with substantial SOC are the 
multilayers
$(\text{SrTiO}_3)_7/(\text{SrIrO}_3)_2$
and $(\text{KTaO}_3)_7/(\text{KPtO}_3)_2$ formed by perovskites
grown along the (111) direction.

First the structure is optimized for different $c$ lattice parameters
respecting IS using GGA and without SOC. This means that
mainly only the
inter-planar distances in the multilayers are relaxed.
For the energy minimum the band structure is calculated turning 
on SOC.

The band structure using three exchange-correlation schemes
(GGA, LDA+U and TB-mBJ) develops the same structure.
The $\nu_b$ topological invariant of each band is calculated as in the TB model,
\cite{prb-kane} the topological invariant being the sum of
the $\nu$ invariants over
all
the occupied bands.
Figure 
3
shows the band structure calculated with TB-mBJ as well as
the $\nu_b$ invariants also obtained ab initio. The difference in the
curvature of the bands with respect to the result obtained with the TB model
is due to the existence of bands near the bottom of the j=3/2 $t_{2g}$
quadruplet which are not considered in the TB Hamiltonian.
Each band has double degeneracy due to the combination of TR and IS.
At the optimized c, the gap between the last filled
and the first unfilled band is located at the corner of the
Brillouin zone (K-point).
At low c (unstable energetically but attainable
via uniaxial compression), GGA predicts that the system can
become a metal by closing
an indirect gap between the K and M points, however TB-mBJ calculations
predict that a direct gap is localized at the K point.
For all
the calculations the ground state has $\nu=1$ and thus it
develops a topological phase. The last filled band has $\nu_b=0$
so by comparison with the TB results the system
corresponds to the
intermediate SOC limit
in which the $J_{eff}=1/2$ and $J_{eff}=3/2$ are not
completely decoupled.\cite{1evir} If a 5d electron system like this is in the intermediate SOC limit, it is hard to imagine how one can build a TMO heterostructe closer
to the strong SOC limit (the only simple solution
would be to weaken the hopping between the TM somehow to increase the 
$\alpha/t$ ratio). The first unfilled band has $\nu_b=1$ as
expected since a $t_{2g}^6$ configuration will be a trivial insulator with a gap
opened by the octahedral crystal field.

\begin{figure}

                \includegraphics[width=\columnwidth]{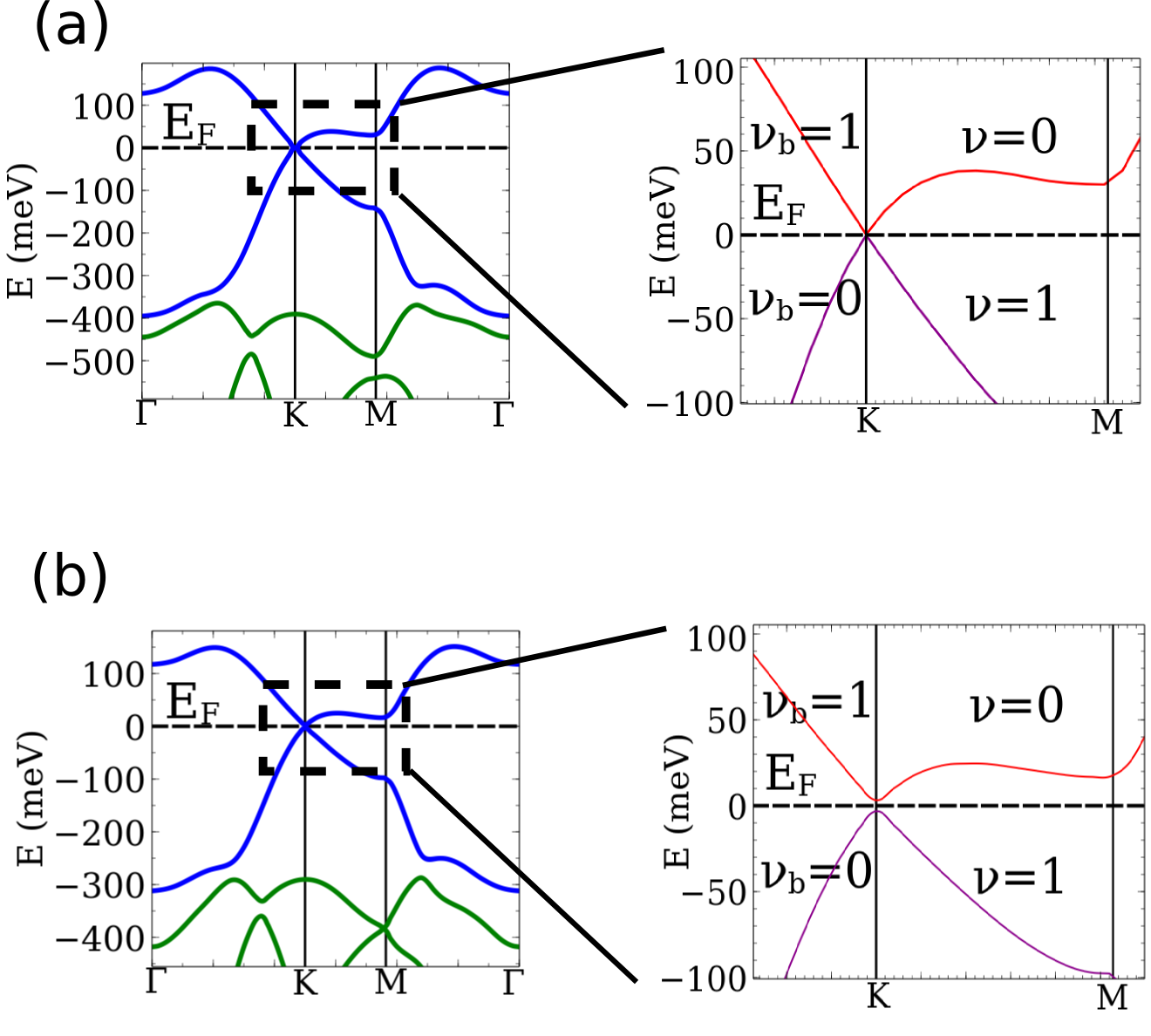}



\caption{(Color online) Band structure obtained in the DFT calculations
for the optimized lattice parameter c. The calculations
were performed with TB-mBJ for both
$(\text{SrTiO}_3)_7/(\text{SrIrO}_3)_2$
(a)
and $(\text{KTaO}_3)_7/(\text{KPtO}_3)_2$
(b). The right panels are the band structure zoomed in near the Fermi level with the
topological invariants displayed. $\nu_b$ is the invariant of the band
considered calculated by Eq. \ref{parities} whereas $\nu$ is the 
sum over all the bands up to the one considered,  in Eq.
\ref{par-full}. }
\label{bandsdft}
\end{figure}

The way a trigonal field is present in the DFT calculations is mainly
in two ways.
On one hand,
 strain along the z direction varies the distance
to the first neighbors in that direction,
so that the electronic repulsion varies as well.
We define this deformation as $\epsilon_{zz}=\frac{c-c_0}{c_0}$ where
$c_0$ is the off-plane lattice constant with lowest energy.
On the other hand, an on-site
Coulomb repulsion
defined on the TM
by using the LDA+U method
has precisely the symmetry of the bilayer, i.e. trigonal
symmetry, so varying in some way the on-site potential (always preserving
parity symmetry) will have the effect of a trigonal term in the
Hamiltonian (see A.3 for further details).

According to this, it is
expected that in a certain regime, variations in $\epsilon_{zz}$
can be compensated by tuning U. In this regime, similar to what we discussed
above, the system will
develop a transition between two topological phases: a 
LUTI and a HUTI. At
even higher U the system will show magnetic order.
We will address this point later and by now we will focus first on the non-magnetic
(NM) phase.
In order to study the phase diagram defined by the parameters $\epsilon_{zz}$ 
and U, we will perform calculations keeping one of them constant
and determine how the gap closes as the other parameter varies, keeping track of
the parities at both sides of
the transition. 

\begin{figure*}


                  \includegraphics[width=2.0\columnwidth]{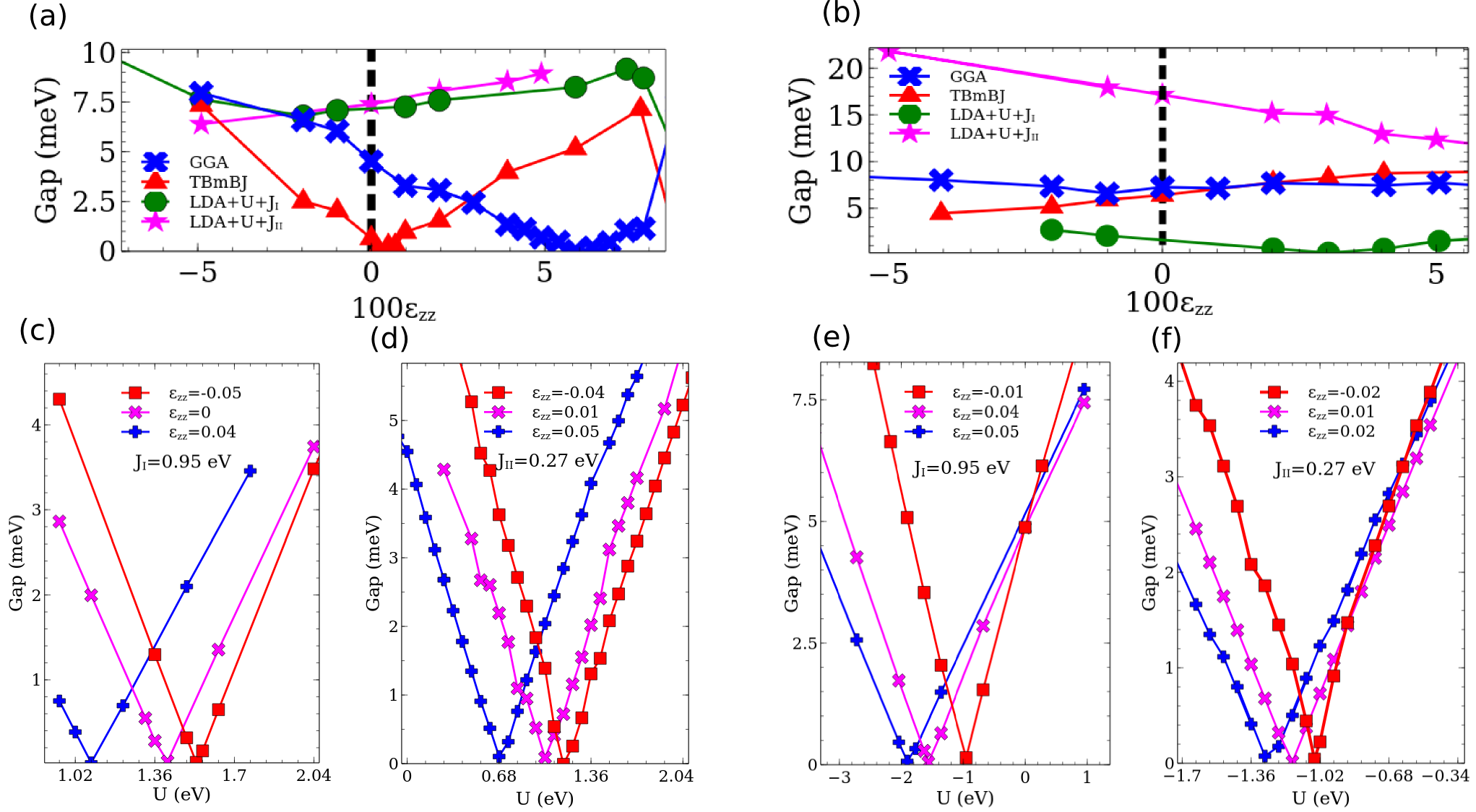}




\caption{(Color online) Evolution of the gap as a function of the uniaxial strain
for the Ir-based multilayer (a) and Pt-based multilayer (b). Evolution of the
gap with the on-site interaction for the Ir-based multilayer (c,d) and Pt-based
multilayer
(e,f) for small (c,e)  and large (d,f) J. It is seen that both systems show a similar behavior although in a
different U regime. In (a) the values of U,J are
$U=2.7$ eV $J=0.95$ eV for the circles and $U=2.3$ eV $J=0.27$ eV
for the stars, in (b) $U=-1.4$ eV $J=0.95$ eV for the circles and $U=2.3$ eV $J=0.27$ eV for the stars. }
\label{evolc}
\label{evolu}
\end{figure*}

\subsection{Evolution of the gap with uniaxial strain}
Here we will discuss the behavior of the gap in the K point with $\epsilon_{zz}$
for various U and J
values. First we analyze the behavior of both materials in
parameter space showing their similarities finishing characterizing the
actual position of the ground state of the system in the general phase
diagram.

First we focus on the $(\text{SrTiO}_3)_7/(\text{SrIrO}_3)_2$ case.
As shown in Figure 4a,
the gap closes as a function
of c, so uniaxial strain can drive the system between two insulating phases
just as $\lambda_{tri}$ does in the TB model.
However, for high U
(see below the discussion on the plausible U values)
the transition point disappears
(two such cases are plot in Fig. 4a).
For low U (see Figs. 4 c,d), $\epsilon_{zz}$ can drive the system from
a positive trigonal term to a negative one. This means that uniaxial
strain can change the sign of the effective trigonal term of the Hamiltonian.
However, for high U  (Fig. 4a),
strain
is not capable of changing the trigonal field,
so the system remains in the same topological phase
for every $\epsilon_{zz}$.
For the calculations using the TB-mBJ scheme (we will see below to what
effective U this situation would correspond),
the transition point takes place almost
at $\epsilon_{zz}=0$, so based on this scheme,
the Ir-based multilayer would be classified rather as
a TSM than as a TI.

Now we will focus on the $(\text{KTaO}_3)_7/(\text{KPtO}_3)_2$ system
(see Fig. 4 b,e,f).
For the GGA calculations the transition with $\epsilon_{zz}$ disappears,
being the system always in the
HUTI phase. If the system is
calculated using LDA+U with U negative
(circles in Fig. 4b),
the behavior is similar to the
previous system and the transition point across a TSM
reappears. For more realistic values of U and J
(stars in Fig. 4b) the transition point disappears again.
Thus, the present system
(Pt-based), though isoelectronic and isostructural, can
be understood as the strong-U limit of the previous system (Ir-based). The band gap is
larger,
which is a sought-after feature of these TI,
but not large enough
to make it suitable for room temperature applications. We observe that 
changing J does not vary the overall picture, just displaces slightly 
the phase diagram in U-space.

\subsection{Evolution of the gap with U at constant $\epsilon_{zz}$}
The Hamiltonian felt by the electrons depends also on the
on-site Coulomb interaction between them.
If the variation in the term of the Hamiltonian that controls it
takes place only in the TM atoms, the
symmetry of the varying term will have the same local symmetry as the TM
atoms, i.e. trigonal symmetry. Thus, it is expected that a variation
in U will have a similar effect as $\lambda_{tri}$ in the TB model, so
the gap can also be tuned by the on-site interaction.

Figure \ref{evolu} c,d,e,f
shows the behavior of the gap for both systems
in an LDA+U scheme
with J=0.27 (realistic) and 0.95 (large) eV
as the parameter U is varied.

The slow increase of the gap with U
suggests that the gap opened is not that of
a usual Mott insulator and reminds
rather to the slow increase obtained in the TB model. 
In fact, the calculation of the
$\nu$ invariant shows again a topological phase on both sides of the transition.

Both systems develop a transition between the LUTI
to the HUTI by increasing U. However, the transition
point of $(\text{SrTiO}_3)_7/(\text{SrIrO}_3)_2$ is at positive U's,
for the $(\text{KTaO}_3)_7/(\text{KPtO}_3)_2$ the
transition appears at negative U's, so for all possible reasonable U values
the system will be in the HUTI phase.

TB-mBJ calculations have proven to give accurate results
of band gaps in
various systems,\cite{test_mbj,test_mbj2,test-mbj-blaha,prb-antia}
including s-p semiconductors, correlated insulators and d systems,
however it might give an inaccurate position of 
semicore d orbitals\cite{test_mbj2}
and overestimate magnetic moments for
ferromagnetic metals.\cite{test-mbj-blaha}
For $(\text{SrTiO}_3)_7/(\text{SrIrO}_3)_2$ it is possible
to use the transition between the LUTI and the HUTI
with the TB-mBJ scheme
to estimate which value of U should be used in an
LDA+U calculation for these 5d systems to reproduce the result
of the TB-mBJ calculation. The actual value of U needed for a correct
prediction of the properties under study is often a matter of contention
when dealing with insulating oxides containing 5d TM's.\cite{Na2IrO3,piro-ir}
From Fig. 4 c,d it
can be checked that the gap closes at U=1.4 eV for
J=0.95 eV and at U=1.0 eV for the more realistic J=0.27 eV, so this
suggests that the values which might be used in an LDA+U
calculation to mimic the TB-mBJ result (a zero gap at $\epsilon_{zz}=0$)
are on the order of $U=1.0$ eV in agreement with Ref. \onlinecite{1evir}.
Other works relate to a value of
U on the order of 2 eV, \cite{martins,arita} however due to the well known
property dependence of the value of U, \cite{u-ceo} it is still unclear
which is the correct value to study these topological phases.
Therefore, our result
can serve as a reference
for other ab initio based phase diagrams for iridates where topological phases
have been predicted as a function of U.\cite{piro-ir} Moreover, we study
this system in a broad range of U values and using different exchange-correlation
schemes to provide a broad picture of the system,
rather than using a fixed U value that would 
yield a more restricted view of the problem.

For the Pt-based multilayer
we could also consider
the hypothetical effective value for which gap would close
at negative U
($U_{eff}=U-J$) for both values of $J$, we obtain the values
$U_{eff}=-1.42$ eV for
$J=0.27$ eV and $U_{eff}=-2.1$ eV for $J=0.95$.
So, it is clearly seen that the gap of these systems is not
only dependent on the
parameter $U-J$, but also has an strong dependence on $J$, both
in the realistic picture of the Ir-based multilayer as well as in
the negative U regime of the Pt-based multilayer.

Recently topological phases
dominated by interactions, called topological Mott insulating phases,
have been theoretically found,\cite{top-mot-ins,ir-top-mot-ins}
being this term employed for physically different phenomena.
In the HUTI phase, the
topological gap of the systems is enhanced by increasing the U parameter
so that
the system seems to be robust against electron-electron interactions.
In the same fashion a usual band insulator can
be connected to a Mott insulating phase,\cite{band-mot,band-mot2}
the previous robustness suggests
that the HUTI phase might be adiabatically connected to a
$Z_2$ non-trivial interacting topological phase.\cite{z2-inter1,z2-inter2}
Whether this is an artifact of the DFT method
or an acceptable mean field approach of a many body problem is something
that can only be clarified with experiments.

\section{Stability of the topological phase}
So far we have considered a system with both TR and IS. However, given that
the $Z_2$ classification
is valid only for TR invariant systems it is necessary to determine
if the ground state possesses this symmetry.
IS breaking could destroy the topological
phase opening a trivial gap by decoupling both sublattices,
as would happen if
the honeycomb is sandwiched by two different materials.\cite{xiao}
TR symmetry breaking
will be fulfilled by a magnetic ground state whereas IS breaking
will be realized by a structural instability. In this Section we will
study the two
possibilities and conclude that both systems are
structurally stable and remain NM according to TB-mBJ
in their ground states.

\subsection{Stability of time reversal symmetry}

\begin{figure}


                \includegraphics[width=\columnwidth]{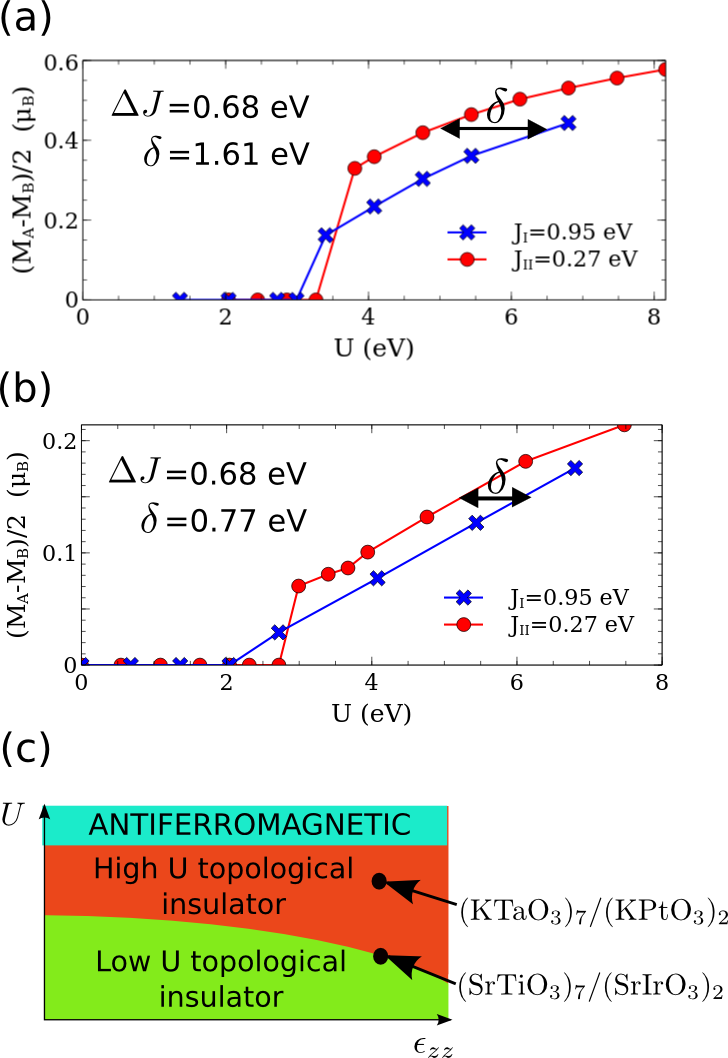}



\caption{(Color online) Difference of the sublattice magnetization for the
$(\text{SrTiO}_3)_7/(\text{SrIrO}_3)_2$
(a) and 
$(\text{KTaO}_3)_7/(\text{KPtO}_3)_2$
(b) as a function of U for two different J.
(c) Phase diagram in $\epsilon_{zz}$-U space,
the approximate position
of the ground state
of both systems according to TB-mBJ is indicated.
At high U the systems develop
an AF order, however at realistic U's both systems remain non-magnetic.}
\label{nm-af}
\end{figure}

\begin{figure*}


                  \includegraphics[width=2.0\columnwidth]{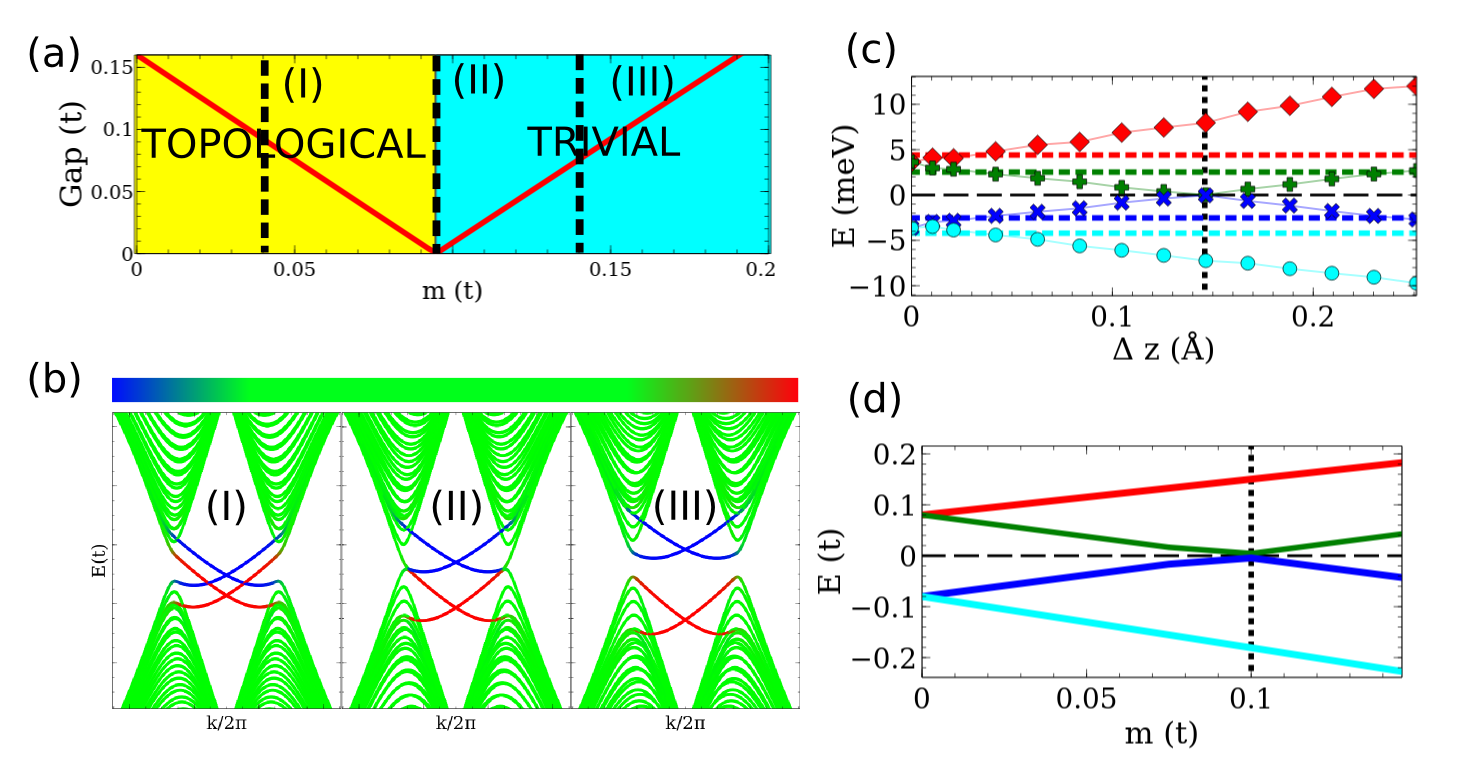}



\caption{(Color online) (a) Evolution of the gap in the TB model
as a function of $m$, the closing point
marks the transition between topological and trivial phase.
(b) Bands of zig-zag ribbons at different $m$,
for low $m$ there are gapless edge states while for large $m$ the edge states become
gapped. The color marks the position of the wavefunction,
the left edge (blue), the right edge (red) or the bulk (green).
(c) Evolution of the eigenvalues obtained in the DFT calculation
(using the GGA scheme) of
 $(\text{KTaO}_3)_7/(\text{KPtO}_3)_2$, the dashed lines correspond
to the fully relaxed structure.
(d) Evolution of the four eigenvalues closest to the Fermi level for
the TB model.
}
\label{break_par}
\end{figure*}

Increasing electronic interactions will drive the NM
ground state to a magnetic trivial Mott
insulating phase at very high U.
For a magnetic $d^5$ S=1/2 localized-electron system,
from Goodenough's rules\cite{goody} an antiferromagnetic (AF)
exchange between the two sublattices is expected which would create an
AF ground state breaking both TR and IS.
To check this result, we have performed DFT calculations within an LDA+U scheme
taking $J_I=0.95$ eV and $J_{II}=0.27$ eV and varying U.
For both systems the calculations have been
carried out at different U's for the NM
and AF configurations at the two J's. In both cases the ground state is
AF at high U. Also,
the sublattice magnetization increases with U.
Figures 5a and 5b
show the evolution of the sublattice magnetic moment
for both compounds. 

Also, a ferromagnetic (FM) phase has been analyzed, being
the least preferred one.
In the Ir compound a FM phase can be stabilized but
has always higher energy than the NM or AF. In the Pt compound a FM
phase could not be stabilized for any of the U values considered.
The true ground state of the system would be a TI phase in the
Pt-based multilayer (or TSM in the Ir-based multilayer according to TBmBJ) depending on the
value of U employed, so if the correct value to be used is larger
than the critical value (of about 3 eV, which is large according to our previous discussions), the topological $Z_2$ phase will break down
and the Kramers protection of the gapless edge states will disappear;
whether the edge states would become gapped or not requires
further study. Thus the experimental
measurement of the magnetic moment of the ground state of these bilayers
would shed light
into the correct value of U which should be used in these and
other similar compounds.
In the Pt-based multilayer the sublattice magnetization is almost only
dependent on $U_{eff}$ as can be checked by the shifting of the curves (Fig. 5b),
however in the Ir-based multilayer there
is a stronger dependence on J (Fig. 5a).
Again, simplifying the evolution in terms of the effective $U_{eff} =U -J$
is discouraged for these systems according to our results.
The non trivial effect of J\cite{review-ldau}
has been also observed in several
compounds such as multi-band
materials.\cite{luca,luca2}

To summarize, we have obtained the magnetization of both compounds as a function
of U, showing
the system
shows a NM ground state for both compounds until a certain U (larger
than the value of U that would be equivalent to TB-mBJ calculations) where
the system becomes AF (see Fig. 5c).

\subsection{Stability of inversion symmetry}
The topological properties of this system rely on 
both TR and IS.
Tight-binding calculations predicted that non-invariant parity terms
with energy associated of the order of magnitude of the topological gap
could eventually drive the topological phase to a trivial one. 

First we will discuss the $(\text{SrTiO}_3)_7/(\text{SrIrO}_3)_2$ system.
The simplest IS breaking could be driven by a structural instability.
To study the structural stability,
we have displaced
one of the TM atoms from its symmetric position and then relaxed the
structure. As a result, the structure returned to
the symmetric configuration. However, due to being in the transition point
between the two topological phases, any external perturbation (such as
a perpendicular
electric field) could break inversion symmetry. This system should be classified
more as a topological semimetal rather than a topological insulator due
to the (almost) vanishing gap of the relaxed structure.

Now we will proceed to $(\text{KTaO}_3)_7/(\text{KPtO}_3)_2$.
The first difference between this and the previous system is that in
the present case the structure is well immersed in the HUTI phase.
A large IS breaking is expected to drive the system to
a trivial phase where the sublattices
A and B would be decoupled. However, to change its
topological class, the system has to cross
a critical point where the gap vanishes in some point of the Brillouin zone.
According to that, the expected behavior is that the gap closes and
reopens as the sublattice asymmetry grows, going from the original
topological phase to a trivial insulating phase.

To check this, we will compare the results obtained from the TB model and DFT
calculations.
In the TB case, we introduce a new parameter which is a
diagonal on-site energy whose value is $+m$ for A atoms and $-m$ for B atoms. 
This new parameter will break IS and its value will take into account the
amount of breaking. When IS is broken the
index $\nu$ cannot be calculated with the parities at the TRIM's. However, we
can study
the topological character searching for gapless edge states. For that sake we
calculate the band structure of a zig-zag ribbon of 40 dimers width with
$\alpha=t$ and $\lambda_{tri}=-t$.
The calculations were also carried out in an armchair ribbon and the same
behavior is found (not shown), however the mixing of valleys
makes the band structure harder to understand.
The color of the bands indicates the expectation value of the
position along the width of the
ribbon of the eigenfunction corresponding to that eigenvalue, it is checked
that the edge states are located in the two edges (red and blue)
whereas the rest of the states are bulk states (green). 
According to the result of the
TB model shown in Figs. 6a
and 6b,
for small $m$ the system remains
a topological insulator although the introduction of $m$ weakens the gap.
If $m$ keeps increasing the system reaches a critical
point where the gap vanishes
and if $m$ increases even more the gap reopens but now the edge states
become gapped so the system is in a trivial insulating phase.

To model
the symmetry breaking in the DFT calculations we move
one of the
Pt atoms in the z direction. As the distance to the original
point increases IS gets more broken.
We show
the four closest eigenvalues to the Fermi level in the K point
for the DFT
calculations (Fig. 
6c) and TB model (Fig.
6d).
For the symmetric structure ($\Delta z=0$), the
combination of TR and IS guarantees
that each eigenvalue is two-fold degenerate. Once the atom is moved the
degeneracy is broken and the eigenvalues evolve with the IS breaking
parameter.
The analogy between the two calculations suggests that in
the DFT calculation once the gap reopens the new state is
also
a trivial insulator. The dashed lines in Fig. 
6c
correspond
to the eigenvalues at the K point for the fully relaxed structure allowing
IS breaking. Comparing with the curves obtained for the evolution
of the eigenvalues it is observed that the relaxed structure is in a slightly
asymmetric configuration but it remains in the HUTI
phase.

Due to the dependence of the topological state on IS, tuning 
this behavior would allow to make a device formed by a perovskite
heterostructure
which can be driven from a topological phase to a trivial phase
applying a perpendicular electric field. The device will be formed by the
TM honeycomb lattice sandwiched between layers of the 
same insulating (111) perovskite from above and below. In this configuration the
system will be in a HUTI phase. However, a perpendicular
electric field will break more the
sublattice symmetry inducing a much greater mass term
in the Hamiltonian proportional to the applied field. Modifying the value of the
electric field it would be possible to drive the system from the
topological phase ($\vec E=0$), to the trivial phase (high $\vec E$). This
can be exploited as an application of this TI in nanoelectronic and spintronic
devices.
\cite{device1,device2,device3,device4} The
sublattice asymmetry needed to make the transition is of the
same order of magnitude as the topological gap, as can be checked in Fig.
6a.

\section{Summary}

We have studied the gap evolution in the $t_{2g}^5$ perovskite
multilayers
$(\text{SrTiO}_3)_7/(\text{SrIrO}_3)_2$ and
$(\text{KTaO}_3)_7/(\text{KPtO}_3)_2$ as a function of the on-site Coulomb
interactions and uniaxial perpendicular strain conserving time
reversal and inversion symmetry. The behavior of the
system has been understood with a simple TB model where SOC
gave rise to an effective j=1/2 four-band Hamiltonian. Uniaxial strain and
on-site interactions have been identified as a trigonal term in the TB model
whose strength controls the magnitude of a topological gap.
The topological invariant $\nu$ has been calculated
using the parities at the TRIM's both in TB and DFT calculations. Comparisons
between the invariants of the bands determines that both
of these 5d electron systems stay in the
intermediate SOC regime.
The small value of
the gap in the K-point comes from being a contribution of third order
in perturbation theory. In contrast,
sublattice asymmetry
contributes as a first order term,
so the topological phase can be easily destroyed by an external perturbation
that gives rise to an IS-breaking term in the Hamiltonian.

$(\text{SrTiO}_3)_7/(\text{SrIrO}_3)_2$ has been found to be a topological
semimetal at equilibrium $\epsilon_{zz}$. Comparing TB-mBJ and LDA+U
calculations 
reasonable results can be obtained for U in the range 1 - 2 eV.
In $(\text{KTaO}_3)_7/(\text{KPtO}_3)_2$ a
HUTI phase at equilibrium has
been found. This last system can be driven
from topological insulating state
to a trivial one by switching on a perpendicular
electric field which would break inversion symmetry.
Also, we have verified
that the properties of these systems are
dependent on both $U$ and $J$ instead of
only in $U_{eff}=U-J$.

Although the smallness
of the gap
(less than 10 meV according to TB-mBJ)
makes the $t_{2g}^5$ configuration
not particularly attractive for
technological applications, the simple understanding of the system turns it
physically very interesting. The present system
can be thought of as an adiabatic deformation of a mathematical
realization of the four band graphene with SOC, with an experimentally
accessible gap, the roles of
$\vec S$ and
$H_{SO}$ being
played now by
$\vec J_{eff}$ and
$H_{tri}$, but with a different physical nature of the topological
state.

\section*{Acknowledgments}
The authors thank financial support from the Spanish Government via the project
MAT-200908165, and V. P. through the Ram\'on y Cajal Program. We also thank
W. E. Pickett for fruitful discussions.

\appendix

\section{Tight binding model}
In this Section we explain the form of the different terms
of the TB Hamiltonian
\begin{equation}
H=H_{SO}+H_t+H_{tri}+H_m
\end{equation}

\subsection{Spin-orbit term}
We want to obtain the form of the SOC restricted to the $t_{2g}$ subspace.
Taking the basis
\begin{equation}
|yz\rangle=
 \begin{pmatrix}
1 \\
0 \\
0
 \end{pmatrix}
\quad
|xz\rangle=
 \begin{pmatrix}
0 \\
1 \\
0
 \end{pmatrix}
\quad
|xy\rangle=
 \begin{pmatrix}
0 \\
0 \\
1
 \end{pmatrix}
\end{equation}

can be easily seen that the representation $L_i=\hbar l_i$ takes the form
\begin{equation}
l_x=
 \begin{pmatrix}
0&0&0 \\
0&0&i \\
0&-i&0
 \end{pmatrix}
\quad
l_y=
 \begin{pmatrix}
0&0&i \\
0&0&0 \\
-i&0&0
 \end{pmatrix}
\end{equation}
\begin{equation*}
l_z=
 \begin{pmatrix}
0&i&0 \\
-i&0&0 \\
0&0&0
 \end{pmatrix}
\end{equation*}

The SOC term has the usual form
\begin{equation}
H_{SO}
=\frac{2\alpha}{\hbar^2}\vec L \cdot \vec S
=\alpha \vec l \cdot\vec\sigma
\end{equation}

where $\sigma_i$ are the usual Pauli matrices acting on spin space.
The representation follows the commutation relation $[l_i,l_j]=-i\epsilon_{ijk}l_k$
so defining $J_i=S_i-L_i$ the usual commutation relations hold.
This change of sign introduces
an overall minus sign on the eigenvalues so the spectrum results in a
$j=3/2$ quadruplet with $E_{j=3/2}=-\alpha$ and $j=1/2$ doublet with
$E_{j=1/2}=2\alpha$.

\subsection{Hopping term}
Each site has three bonds directed along the edges of the perovskite
structure.
The bonds link the A and B sublattices so
there will be three different overlaps $t_i=\langle A |H_t|B \rangle$ depending
on the direction

\begin{equation}
(t_x,t_y,t_z)=t(\tau_x,\tau_y,\tau_z)
\end{equation}

The hopping takes place through the overlap of the $t_{2g}$ orbitals of the
TM and the oxygen p orbitals,
It can be easily checked
that the main contribution gives a matrix form
that can be casted in the form
\begin{equation} 
(\tau_x,\tau_y,\tau_z)=(l_x^2,l_y^2,l_z^2)
\end{equation}

\subsection{Trigonal term}
Due to the geometry of the system, a possible term in the Hamiltonian
that does not break the spatial symmetries of the system is
a trigonal term. This term will differentiate the perpendicular
direction from the in-plane directions. This term will behave as an on-site
term which mixes the $t_{2g}$ states without breaking the symmetry between them
so the general form in the $t_{2g}$ basis is 

\begin{equation}
H_{tri}=\lambda_{tri}
 \begin{pmatrix}
0&1&1 \\
1&0&1\\
1&1&0
 \end{pmatrix}
\end{equation}
given that the previous matrix is diagonal in the (111) direction,
being the perpendicular eigenvalues degenerated, thus preserving
the trigonal symmetry.

The way $\epsilon_{zz}$ enters in this term is
on one hand by anisotropy of charge density due to the lack of local
spatial inversion and on the other hand by distortion of the cubic
perovskite edges (and thus of the octahedral environment)
by expansion/contraction of the (111) direction. The absence of local 
octahedral rotational
symmetry is also responsible for the dependence of $\lambda_{tri}$ on U.
Since varying the onsite interaction will modify the local electron density,
provided the local trigonal symmetry is conserved but not the local inversion
(as it is broken explicitly by the multilayer),
this modification in the electronic density will influence the electrons
across the Hartree and exchange-correlation
terms, by a term with those symmetries.
In summary, local symmetry forces that
a spin-independent perturbation, that includes $\epsilon_{zz}$
and spin-independent U-terms, can be recasted on the previous form.

Whether spin-mixing trigonal
terms are relevant for the effective model of this system
or not
should be clarified in the future. Either way, the agreement of the
predictions both of the TB model and the DFT calculations, in addition with
the explicitly checked topological phase at high U suggests that this model
is a well behaved effective model to study the NM phase of
this type of systems.

\subsection{Mass term}
A term which makes the two sublattices nonequivalent will break inversion
symmetry. The minimal term which fulfills this is 

\begin{equation}
H_{m}=ms_z=m
 \begin{pmatrix}
I_A&0 \\
0&-I_B\\
 \end{pmatrix}
\end{equation}

where $I_A$ and $I_B$ are the identity matrix over the A and B sublattices


\end{document}